# Efficient Construction of Dominating Set in Wireless Networks

Xiaohua Xu
Department of Computer Science
Illinois Institute of Technology
Chicago, IL 60616
Email: xxu23@iit.edu

Xiang-Yang Li
Department of Computer Science
Illinois Institute of Technology
Chicago, IL 60616
Email: xli@cs.iit.edu

*Abstract*—Considering a communication topology of a wireless network modeled by a graph where an edge exists between two nodes if they are within each other's communication range. A subset $U$ of nodes is a dominating set if each node is either in $U$ or adjacent to some node in $U$. Assume each node has a disparate communication range and is associated with a positive weight, we present a randomized algorithm to find a min-weight dominating set. Considering any orientation of the graph where an arc $\overrightarrow{uv}$ exists if the node $v$ lies in $u$'s communication range. A subset $U$ of nodes is a strongly dominating set if every node except $U$ has both in-neighbor(s) and out-neighbor(s) in $U$. We present a polynomial-time algorithm to find a strongly dominating set of size at most $(2 + \epsilon)$ times of the optimum. We also investigate another related problem called $K$-Coverage. Given are a set $\mathcal{D}$ of disks with positive weight and a set $\mathcal{P}$ of nodes. Assume all input nodes lie below a horizontal line $l$ and all input disks lie above this line $l$ in the plane. The objective is to find a min-weight subset $\mathcal{D}' \subseteq \mathcal{D}$ of disks such that each node is covered at least $K$ disks in $\mathcal{D}'$. We propose a novel two-approximation algorithm for this problem.

## I. INTRODUCTION

Dominating set can serve as a backbone of wireless networks and has has a wide range of applications. Many kinds of activities for a wireless networking system, such as routing, broadcasting, and topology control rely on dominating set. From both economic and practically applicable concerns, we require a dominating set in a wireless networking system to sustain as long as possible. However, most wireless nodes are powered by batteries and have a stringent energy budget. We will address this challenge and design efficient schemes to achieve energy efficiency. Despite of those applications, dominating set is a classical problem and has drawn a lot of research interest in computational geometry independently.

Given a wireless network in an Euclidean plane, its communication topology is modeled by a graph where there exists an edge between two nodes if and only if they are within each other's the communication ranges. This graph is called *disk containment graph* in the sense that the communication range of each node is a disk. If we represent each node $u$ as a disk $D_u$ centering at $u$ with radius equal to the transmission range of node $u$, we then connect each pair of disks if both disks contain the centers of the other one, *i.e.*, the corresponding two nodes $u$ and $v$ are within the transmission ranges of each other. A subset $U$ of nodes is a dominating set if each node is either in $U$ or adjacent to a node in $U$. Assume each node is associated with a positive weight, the problem **Minimum Weight Dominating Set (MWDS)** seeks a dominating set of minimum total weight. This problem has been studied extensively in multi-hop wireless networks with uniform communication ranges. However, in practice the nodes may have different communication ranges either because of the heterogeneity of the nodes, or due to interference mitigation, or due to a chosen range assignment for energy conservation. There has also been extensive work for dominating set problem in the setting of *disk intersection graph* where there exists an edge between two nodes if their communication disks intersect with each other.

We then consider dominating set in directed graph. Given a wireless networking system represented by a digraph $G = (V, E)$. Each node $u$ is represented as a disk $D_u$ centering at $u$ with radius equal to the transmission range of node $u$. We then draw a directed edge $\overrightarrow{uv}$ if the disk $D_u$ contain the center of $D_v$, *i.e.*, the corresponding nodes $u$ and $v$ are within the transmission ranges of each other. A subset of nodes $U \subseteq V$ is a *strongly dominating set* if every node in $V \setminus U$ has both an in-neighbor in $U$ and an out-neighbor in $U$. The problem **Minimum Strongly Dominating Set (MSDS)** seeks a strongly dominating set of minimum cardinality. We will reduce this problem to *Minimum Disk Cover* and *Geometric Hitting Set*, and apply their algorithmic results to solve it.

The motivation for studying MSDS is described as follows. Given a wireless network, to construct a backbone for routing and broadcasting, we need to ensure that for every node $v$, there exist a node $u$ such that both $u$ and $v$ are within the transmission ranges of each other, thus node $v$ can both send and receive data from the network. Traditional, this is modeled as selecting a dominating set in disk containment graph (*i.e.*, the MWDS problem). Note that, to achieve the purpose, we actually can relax the requirement to that: for any node $v$, there exists a node $u_1 \in U$ such that $v$ can transmit its data to $u_1$ ($u_1$ is within of the transmission range of $v$), and at the same time, there exists a node $u_2 \in U$ such that $v$ can be receive data from $u_2$ ($v$ is within of the transmission range of $u_2$). These nodes $u_1$ and $u_2$ are not necessarily the same. This is actually to select a strongly dominating set in a redirected version of disk containment graph. A strongly dominating set can ensure that for any node $v$, there exists two nodes $u_1, u_2 \in U$ such that $v$ can transmit its data to $u_1$ and $v$ can be receive data from $u_2$

Since dominating set is intrinsically related to coverage, we also investigate a classical disk cover problem. Suppose that we are given with a set $\mathcal{D}$ of disks with positive cost (or weight) defined by a function $c : \mathcal{D} \mapsto N^+$ and a set $\mathcal{P}$ of nodes in the plane. A node $p \in \mathcal{P}$ is covered by a disk $D \in \mathcal{D}$ if $p$ lies in $D$. A subset $\mathcal{D}' \subseteq \mathcal{D}$ is said to be a $K$-cover of $\mathcal{P}$ if each node in $\mathcal{P}$ is covered by at least $K$ disks in $\mathcal{D}'$. Assume all input nodes lie below a horizontal line $l$ and all input disks lie above this line $l$. The problem **Linear K-Cover (LKC)** seeks a $K$-cover $\mathcal{D}' \subseteq \mathcal{D}$ of $\mathcal{P}$ with minimum total weight. The disk cover problem is a geometric set cover problem, where the given sets are defined by disks. It has been proved to be NP-hard even when all disks are unit disks and $K = 1$ [9]. However, the geometric restriction can admit a constant-approximation algorithm, and tremendous work is done for various disk cover problems: [2]–[4], [13]. Among those work, a strong assumption that all targets are required to be covered by only once, while in this work, we study coverage in a general setting, *i.e.*, multiple coverage.

**Our Main Contributions:** In this work, we will design algorithms with theoretical analysis for the dominating set and coverage problems respectively.

- For the problem MWDS, we apply a uniform sampling process technique based a recent breakthrough result [8], [15], and present randomized algorithms to solve them. We prove that we can achieve an approximation ratio of $2^{O(\log^* n)}$, with high probability, where $\log^* n$ is the smallest number of iterated "logarithms" applied to $n$ to yield a constant.
- For the problem MSDS, we present a new polynomial-time $(2 + \epsilon)$-approximation algorithm based on a recent breakthrough result [12]. We also provide a heuristic of further treatment of this problem.
- For the problem LKC, we present the first two-approximation algorithm based on a dynamic programming technique.

The rest of the paper is organized as follows: Section II, III and Section IV are devoted to the presentation of our solutions for the problem MWDS, MSDS, and LKC respectively. In Section V, we conduct a thorough literature review. We conclude our paper and discuss possible future research directions in Section VI.

## II. MIN-WEIGHT DOMINATING SET

Given an instance of the problem MWDS in the disk containment graph $G$: a set $\mathcal{D}$ of $n$ weighted disks, for any disk $D$, let $N(D) \subseteq \mathcal{D}$ denote $D$'s neighboring disks in $G$, let the binary variable $x_D \in \{0, 1\}$ indicate whether the disk $D$ is selected in the solution or not. We assume the weight function for $\mathcal{D}$ is $c : \mathcal{D} \mapsto \mathbb{R}^+$. We can formulate it as an integer linear programming.

$$\min : \sum_{D \in \mathcal{D}} c(D) \cdot x_D, \quad \text{s.t. :}$$
$$\begin{cases} \sum_{A \in N(D)} x_A \geq 1, \forall D \in \mathcal{D} \\ x_D \in \{0, 1\}, \forall D \in \mathcal{D} \end{cases} \quad (1)$$

Here the first set of linear constraints in Equation (1) reflects the fact that the resulted solution (a subset of selected disks) is a dominating set in the disk containment graph $G$. We then relax the requirement such that $x_D : D \in \mathcal{D}$ can be any value in $[0, 1]$, instead of only integers. We consider the following linear programming relaxation for the problem MWDS in the disk containment graph.

$$\min : \sum_{D \in \mathcal{D}} c(D) \cdot x_D, \quad \text{s.t. :}$$
$$\begin{cases} \sum_{A \in N(D)} x_A \geq 1, \forall D \in \mathcal{D} \\ x_D \geq 0, \forall D \in \mathcal{D} \end{cases} \quad (2)$$

The LP relaxation admits a polynomial time optimal solution $\{x_D : D \in \mathcal{D}\}$. We then create a set $\mathcal{D}_0$ of disks as follows: for each disk $D$, we add $\lfloor 2n \cdot x_D \rfloor$ copies of $D$ to $\mathcal{D}_0$. For the special case when $x_D < \frac{1}{2n}$, $\lfloor 2n \cdot x_D \rfloor = 0$, we do not add any copy of $D$ to $\mathcal{D}_0$. Each added copy of the disk $D$ inherits $D$'s original weight. For the set $\mathcal{D}_0$ of disks, we observe two important facts.

*Lemma 1:* The following facts are true:
1) Each disk $D$ has at least $n$ neighbors from $\mathcal{D}_0$ in $G$;
2) $w(\mathcal{D}_0) \leq 2n \cdot \lambda^*$, where $\lambda^*$ is the optimal objective function value for the LP-relaxation in Equation (1).

*Proof:* The proof is available in the appendix. ∎

Next, we iteratively apply the *uniform sampling process* (in Table II) to produce a successively sparse dominating set. For the first iteration, we set the input dominating set as $\mathcal{D}_0$, and the parameter $L_1 = n$. For the $i$-th iteration, we set the input dominating set as $\mathcal{D}_{i-1}$, and the parameter $L_i = \log L_{i-1}$, to obtain an output dominating set $\mathcal{D}_i \subseteq \mathcal{D}_{i-1}$ which is sparser, for $i = 2, 3, \cdots, t$ ($t = \log^* n$). Finally, we output $\mathcal{D}_t$. The details are shown in Table I.

---

**Randomized Algorithm**:

**Input**: a set $\mathcal{D}$ of disks, $c : \mathcal{D} \mapsto \mathbb{R}^+$;
Solve the LP relaxation in Equation (1),
let $\{x_D : D \in \mathcal{D}\}$ be the output;
**for** each disk $D \in \mathcal{D}$
   add $\lfloor 2n \cdot x_D \rfloor$ copies of $D$ with the same weight to $\mathcal{D}_0$;
$t \leftarrow \log^* n$;
**for** $i = 0$ to $t - 1$
   apply uniform sampling process (Table II) on $\mathcal{D}_i$;
   let $\mathcal{D}_{i+1}$ be the output;
   $i++$;
delete the redundant disks from $\mathcal{D}_t$;
**return** $\mathcal{D}_t$.

TABLE I
RANDOMIZED ALGORITHM

---

The *uniform sampling process* here is a probabilistic algorithm that takes an input dominating set $\mathcal{D}$, and a parameter $L$, and outputs a *sparse* dominating set $\mathcal{D}'$, where the probability of each disk being selected is at most $\frac{c \log L}{L}$. In addition, each disk that is $L$-dominated by $\mathcal{D}$ is at least $\log L$-dominated by $\mathcal{D}'$. Here a disk is *L-dominated* if it has exactly $L$ neighboring disks from $\mathcal{D}$ in the disk containment graph $G$. In such case, this disk is also said to be dominated with the *multiplicity L*.

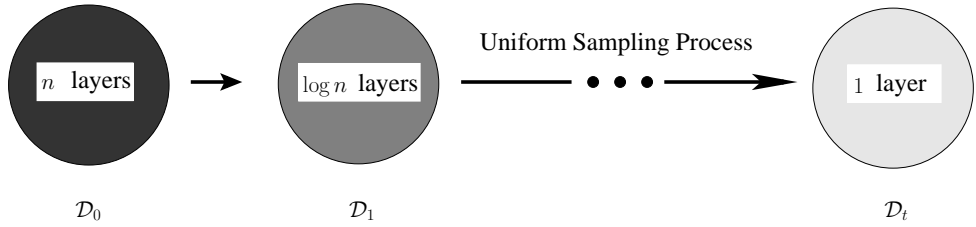

Fig. 1. Illustration for Uniform Sampling Process, the various shades reflect the sparsities of disk sets.

We first only consider the subset $\mathcal{A}$ of disks that are dominated by $\mathcal{D}$ with the multiplicity in $[L, 2L]$ (each disk from $\mathcal{D}'$ is dominated by at least $L$ disks, and at most $2L$ disks from $\mathcal{D}$), we will produce a subset $\mathcal{D}' \subseteq \mathcal{D}$ of disks, such that $\mathcal{A}$ will be at least $\log L$-dominated in $\mathcal{D}'$. We repeat the process for the disks that are dominated by $\mathcal{D}$ with the multiplicity in $[2L, 4L]$, $[4L, 8L]$ and so on, the output is a series of sparse sets of disks. We finally output the union of all disk sets. We will prove that the probability of a disk being selected can still upper bounded by $O\left(\frac{c \log L}{L}\right)$.

**Uniform Sampling Process**:
**Input**: $\mathcal{D}$, $\mathcal{A}$, a parameter $L$;
Construct the sequence $\sigma \leftarrow <D_1, \cdots, D_m>$ on $\mathcal{D}$;
**for** $i = 1$ to $m$
  **for** each disk $A \in \mathcal{A}$ dominated by $D_i$
    **if** $D_i$ is forced because of $A$
      add $D_i$ to $\mathcal{D}'$;
    **else** add $D_i$ with probability $\frac{c \log L}{L}$ to $\mathcal{D}'$;
  $i++$;
**return** $\mathcal{D}'$.

TABLE II
UNIFORM SAMPLING PROCESS

Given two disk sets $\mathcal{A}$ and $\mathcal{D}$, we define an *equivalence class* for $\mathcal{A}$ as a subset of all disk from $\mathcal{A}$ that are dominated exactly by the same set of disks from $\mathcal{D}$. Then, all the equivalence classes for $\mathcal{A}$ induce a partition of $\mathcal{D}$. Note that, if a set $\mathcal{D}'$ at least $\log L$-dominate one disk in an equivalence class, then they at least $\log L$-dominate all disks in that class. Thus, we can assume we have one representative disk from each class. We want to at least $\log L$-dominate these representative disk.

Let $N_m = \mathcal{D}$, and let $C_m$ denote the set of *equivalence classes* of disks such that the disks in each class is dominated with multiplicity at most $2L$. By Lemma 2, $|C_m| \leq c' n_m L^2$, $n_m = |N_m|$. We compute a disk $D_m \in N_m$ that dominates the least number of equivalence classes. By pigeonhole principle, $D_m$ dominates at most $2c'L^3$ classes of $C_m$. We will recursively compute a sequence of disks for a new instance for $N_{m-1} = N_m \setminus \{D_m\}$, and append the sequence to $D_m$. In the new instance for $N_{m-1}$, we consider the classes $C_{m-1}$ whose dominating multiplicity in $N_{m-1}$ is at most $2L$. Let $\sigma$ be the reverse of this sequence: $\sigma = <D_1, \cdots, D_m>$. Note that the method for constructing the sequence $\sigma$ is similar to *smallest last ordering* [10].

We then focus on the sequence $\sigma = <D_1, \cdots, D_m>$. For each disk $D_j$, we decide instantly whether adding it to $\mathcal{D}'$ or not, depending on whether $D_j$ is forced or not. Here we call a disk $D_j \in N_j$ forced if not including $D_j$ will result in a consequence that some equivalence class can not be at least $\log L$-dominated. The details are shown in Table II.

We next analyze the approximation ratio of the proposed algorithm described in Table II. We will make use of the following lemma:

*Lemma 2:* Let $\mathcal{D}$ be a set of $m$ disks, and $1 \leq L \leq m$ be an integer. There are $O(mL^2)$ equivalence classes dominated by distinct subsets of $\mathcal{D}$, each of size at most $L$.

Based on Lemma 2, we can prove Theorem 1, which is a variant of the result in [8], [15].

*Theorem 1:* (**Uniform Sampling Property**) The uniform sampling process (Table II) produces a subset $\mathcal{D}' \subset \mathcal{D}$ out of $\mathcal{D}$, such that for any disk $D$, if $D$ is $L$-dominated in $\mathcal{D}$, then $D$ is at least $\log L$-dominated in $\mathcal{D}'$ and $\Pr(D \in \mathcal{D}') \leq \frac{c \log L}{L}$.

*Proof:* The proof is available in the appendix. ∎

By combining Theorem 1 with Lemma 1, we can obtain the following main theorem.

*Theorem 2:* For the problem MWDS, there exists a randomized algorithm that produces a dominating set $\mathcal{D}_t$, and $c(\mathcal{D}_t) \leq 2^{O(\log^* n)} \cdot c^*$, with high probability, where $c^*$ denotes the weight of an optimal solution, and $\log^* n$ is the smallest number of iterated "logarithms" applied to $n$ to yield a constant.

The proof of Theorem 2 is similar to Section 3.1 in [8].

### A. Discussions on MDS

Let us define a new concept called *restricted dominating set*. Given a disk containment graph $G = (V, E)$, a subset $U \subset V$ is a restricted dominating set if every disk $v \in V$ is either $v \in U$ or $\exists u \in U$ with $r_u \geq r_v$ such that $uv \in E$.

The following theorem reduces the minimum dominating set to its restricted version.

*Theorem 3:* Suppose that there exists a polynomial algorithm for selecting the restricted dominating set of minimum size, then there is polynomial 6-approximation algorithm for selecting the minimum dominating set in a disk containment graph.

*Proof:* Consider an instance $G = (V, E)$, assume $U^*$ is the optimum solution for minimum dominating set, we will construct a restricted dominating set $R$ based on $U^*$ and satisfying: $|R| \leq 6 \cdot |U^*|$. Then it is easy to verify that the optimum solution $R^*$ for minimum restricted dominating set is 6-approximation for the minimum dominating set problem since $|R^*| \leq |R| \leq 6 \cdot |U^*|$.

We next construct $R$ based on $U^*$. First we define a mapping between nodes and disks:

*Definition 1:* For each node $v \in V$, we define its corresponding disk $D_v$ as the disk centering at $v$ and with $r_v$ as the radius.

Observe that all nodes lie inside at least one disk in $\cup_{u \in U^*} D_u$ since $U^*$ is a feasible dominating set. Let $R = U^*$ initially. Thus if we add disks to $R$ to ensure that: for each disk $D_u : u \in U^*$, all nodes inside $D_u$ satisfy the restricted property, then the resulted solution $R$ is a restricted dominating set. We check the disk one by one. For each disk $D_u : u \in U^*$, we will add minimal nodes to $R$ such that all nodes inside $D_u$ satisfy the restricted property. Note that the nodes inside $D_u$ with communication radius no larger than $r_u$ (including $u$) already satisfy the restricted property.

Thus we only need to focus on the nodes $v$ inside disk $D_u$ with $r_v \geq r_u$. Assume the set of nodes is $V_u$. Let $R_u = V_u$. We then continuously delete nodes from $R_u$ as long as each nodes in $V_u$ satisfy the restricted property with respect to $R_u$ after the deletion. Assume the output is $R_u$ after our deletion can not proceed any more. We prove that for each node $u \in U^*$, $|R_u| \leq 5$. Otherwise, there exist two nodes $v, w \in R_u$ such that $\angle vuw \leq 60° \Rightarrow \|vw\| \leq \max\{uv, uw\} \leq r_u \leq \min\{r_v, r_w\}$. Thus, $vw \in E$, then we can delete one node in $\{v, w\}$ with smaller communication range from $R_u$. This contradicts that $R_u$ is the output after our deletion can not proceed any more.

Finally, for each node $u \in U^*$, we add $R_u$ to $R$. Clearly, the resulted $R$ is a feasible solution for restricted minimum dominating set, and $R = U^* \bigcup_{u \in U^*} R_u$, thus $|R| \leq |U^*| + \sum_{u \in U^*} |R_u| \leq |U^*| + \sum_{u \in U^*} 5 \leq 6|U^*|$. ∎

Next, we make a conjecture for selecting restricted dominating set based on dynamic programming technique.

*Conjecture 1:* There exists a constant approximation algorithm for selecting the minimum restricted dominating set.

If the conjecture holds, we can find a constant approximation algorithm for the original problem, *i.e.*, the minimum dominating set problem.

## III. MINIMUM STRONGLY DOMINATING SET

In this section, we present a $(2 + \epsilon)$-approximation algorithm for the problem MSDS.

A forward dominating set is a subset $D$ of $U$ such that for each node $u$, there exist a node $v$ and $\overrightarrow{uv} \in E$. A backward dominating set is a subset $D$ of $U$ such that for each node $u$, there exist a node $v$ and $\overrightarrow{vu} \in E$. Observe that, any strongly dominating set is a union of a forward dominating set and backward dominating set. Our method for selecting a strongly dominating set can be divided into two phases. First, we select a forward dominating set $U_1$. Second, we select a backward dominating set $U_2$. We finally output the union $U_1 \cup U_2$ of two sets (deleting the redundant nodes).

Thus, we have the following main theorem.

*Theorem 4:* Suppose that there exists a polynomial $a$-approximation algorithm for the Minimum Geometric Hitting Set and there exists a polynomial $b$-approximation algorithm for the Minimum Disk Cover, then there is polynomial $(a+b)$-approximation algorithm for the strongly dominating set problem in a disk containment graph.

We observe that the problem forward dominating set selection can be reduced to the problem Minimum Geometric Hitting Set. There exists a polynomial $(1 + \epsilon)$-approximation algorithm for Minimum Geometric Hitting Set [12] by using a Local Search method. Given an instance where $\mathcal{P}$ is the input point set, we begins with the point set $\mathcal{N} = \mathcal{D}$, which clearly is a hitting set. We then replace any point subset with size at most $k$ by a point subset of size at most $k - 1$, if this replacement still results in a hitting set. We will keep replacing until no further possible replacement. The proposed algorithm has been proved to achieve an approximation ratio of $1 + \epsilon$ for the problem Minimum Geometric Hitting Set.

On the other hand, the backward dominating set selection can be reduced to the problem Minimum Disk Cover. For the problem Minimum Disk Cover, there exists a polynomial $(1+\epsilon)$-approximation algorithm [16] by using a Local Search method as well. Given an instance where $\mathcal{D}$ is the input disk set, we begins with the disk set $\mathcal{N} = \mathcal{D}$, which clearly is a cover of nodes $\mathcal{P}$. We then replace any disk subset $\mathcal{B}$ of $\mathcal{N}$ with size at most $k$ by a disk subset of size at most $|\mathcal{B}| - 1$, if the disk set $\mathcal{N}$ after replacement is still a disk cover. We will keep replacing until no further possible replacement. The proposed algorithm has been proved to achieve an approximation ratio of $1 + \epsilon$ for the problem Minimum Disk Cover.

Therefore, we have the following theorem.

*Theorem 5:* The problem MSDS admits a polynomial time $(2 + \epsilon)$-approximation solution.

## IV. LINEAR $K$-COVER

Given an instance of the problem **Linear $K$-Cover**, there are a set $\mathcal{D}$ of disks with positive cost (or weight) given by a function $c : \mathcal{D} \mapsto N^+$ and a set $\mathcal{P}$ of nodes in the plane. We begin with some terms and notations. Note that when involving a disk's location, we refer to the location of the disk's center, later on we will keep this notation. We note all nodes in $\mathcal{P}$ from left to right as $p_1, p_2, \cdots, p_n$. Assume $\mathcal{P}_i$ is the set of nodes from $\mathcal{P}$ lying to the left to $p_i$ (including $p_i$). For the ease of treatment, we also introduce a dummy disk as follows. The low half-plane $y \geq y_1$ defines a dummy disk of zero weight. We denote by $\mathcal{D}^+$ the union of $\mathcal{D}$ and this dummy disk. The dummy disk does not covers any node in $\mathcal{P}$, but it intersects every vertical line as a half-plane.

Consider a disk $D \in \mathcal{D}^+$ intersecting a vertical line $l$. A disk $D' \in \mathcal{D}^+$ is said to be line-dominated by $D$ w.r.t. $l$ if one of following cases occurs.

1) $D'$ does not intersect $l$;

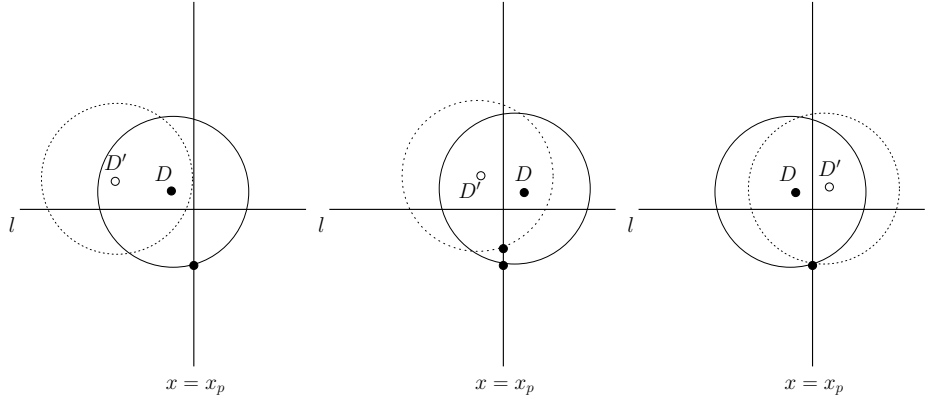

Fig. 2. Three cases when disk $D$ line-dominates $D'$.

2) The lowest endpoint of $D \cap l$ is below the lowest endpoint of $D' \cap l$;
3) $D \cap l$ and $D' \cap l$ have the same lowest endpoint, but $D$ lies to the left of $D'$.

It's easy to verify that *line-dominating* is transitive: Suppose that $D_1, D_2, D_3$ are three disks in $\mathcal{D}_i^+$ for some $2 \leq i \leq m+1$, and $l$ is a vertical line. If $D_1$ line-dominates $D_2$ w.r.t. $l$ and $D_2$ line-dominates $D_3$ w.r.t. $l$, then $D_1$ line-dominates $D_3$ w.r.t. $l$.

*Definition 2:* (skyline) Given a set of disks $\mathcal{D}' \subseteq \mathcal{D}$ and a node $p$, the skyline of $\mathcal{D}'$ at $p$ is a sequence of $K$ disks $D_1, D_2, \cdots, D_K$ from $\mathcal{D}'$ such that $D_K$ line-dominates $D_{k+1}$ at $p$, moreover, $D_K$ line-dominates all disks from $\mathcal{D}' \setminus \{D_1, D_2, \cdots, D_K\}$.

For any node $p_i \in \mathcal{P}$, let $\mathcal{D}_i$ denote the set of disks in $\mathcal{D}$ covering $p_i$. Denote

$$\Gamma_i = \mathcal{D}_1 \times \mathcal{D}_2 \cdots \times \mathcal{D}_K$$

For any $(D_1, \cdots, D_K) \in \Gamma_i$, let $\mathcal{C}_i(D_1, \cdots, D_K)$ denotes the collection of $K$-covers $\mathcal{D}'$ of $\mathcal{P}_i$, which satisfies:

- $(D_1, \cdots, D_K)$ is the skyline of $\mathcal{D}'$ at $p_i$;
- $D_1, \cdots, D_K$ cover $p_i$.

*Definition 3:* ($i$-th startup cost) The $i$-th restart cost of a disk set $\mathcal{D}'$ (noted as $c_i(\mathcal{D}')$) is defined as follows: We proceed the nodes from left to right. For each node $p_i$, we can determine the skyline $\text{SKY}_i$ of $\mathcal{D}'$ at $p_i$. For each disk $D \in \text{SKY}_i$, if it does not appear in $\text{SKY}_{i-1}$, then we add the cost of $D$. Otherwise, we will not add the cost.

If $\mathcal{C}_i(D_1, \cdots, D_K)$ is not empty, let $C_i(D_1, \cdots, D_K) \in \mathcal{C}_i(D_1, \cdots, D_K)$ be a $K$-cover with minimum $i$-th startup cost, and $c_i(D_1, \cdots, D_K)$ be the $i$-th startup cost of $C_i(D_1, \cdots, D_K)$; Otherwise, set $C_i(D_1, \cdots, D_K)$ to null, and set $c_i(D_1, \cdots, D_K)$ to $\infty$.

For any $(D_1, \cdots, D_K) \in \Gamma_i$, we denote by $\Gamma'(D_1, \cdots, D_K)$ the set of $(D'_1, \cdots, D'_K)$ satisfying that,

- $(D_1, \cdots, D_K)$ is the skyline of $\{D_1, \cdots, D_K\} \cup \{D'_1, \cdots, D'_K\}$ at $p_i$.

We prove the following recursive relation. We do not consider the case when $c_i(D_1, \cdots, D_K) = \infty$.

*Theorem 6:* $\forall i, \forall (D_1, \cdots, D_K) \in \Gamma_i$,

$$c_i(D_1, \cdots, D_K) = \min_{(D'_1, \cdots, D'_K) \in \Gamma_{i-1}(D_1, \cdots, D_K)}$$

$$\left\{ c_{i-1}(D'_1, \cdots, D'_K) + c\left( \{D_1, \cdots, D_K\} \setminus \{D'_1, \cdots, D'_K\} \right) \right\}$$

*Proof:* **LHS $\geq$ RHS:** Let $D'_1, \cdots, D'_K$ be the skyline of $C_i(D_1, \cdots, D_K)$ at $p_{i-1}$. Let $\mathcal{D}''$ be the set of disks that appear in any of the previous $i-1$ skylines of $C_i(D_1, \cdots, D_K)$. Set

$$\mathcal{D}' = C_i(D_1, \cdots, D_K) \setminus \left( \{D_1, \cdots, D_K\} \setminus \mathcal{D}'' \right)$$

We have $\mathcal{D}' \in \mathcal{C}_{i-1}(D'_1, \cdots, D'_K)$. Therefore, $c_{i-1}(D'_1, \cdots, D'_K) \leq c_{i-1}(\mathcal{D}')$. In addition, we have

$$c_{i-1}(\mathcal{D}') = c_i(D_1, \cdots, D_K) - c\left( \{D_1, \cdots, D_K\} \setminus \{D'_1, \cdots, D'_K\} \right).$$

Therefore, we have

$$c_i(D_1, \cdots, D_K) \geq c_{i-1}(D'_1, \cdots, D'_K) + c\left( \{D_1, \cdots, D_K\} \setminus \{D'_1, \cdots, D'_K\} \right).$$

**LHS $\leq$ RHS:** Suppose RHS achieves minimum at $(D'_1, \cdots, D'_K)$.

Let

$$\mathcal{D}' = C_{i-1}(D'_1, \cdots, D'_K) \cup \{D_1, \cdots, D_K\}.$$

First, $\mathcal{D}'$ is a $K$-cover of $\mathcal{P}_i$ and $(D_1, \cdots, D_K)$ is the skyline of $\mathcal{D}'$ at $p_i$. Then, $\mathcal{D}' \in \mathcal{C}_i(D_1, \cdots, D_K)$. Consequently, we have

$$c_i(D_1, \cdots, D_K) \leq c_i(\mathcal{D}') \leq c_{i-1}(D'_1, \cdots, D'_K) +$$

$$c\left( \{D_1, \cdots, D_K\} \setminus \{D'_1, \cdots, D'_K\} \right).$$

■

*Theorem 7:* $\min_{D_1, \cdots, D_K} c_n(D_1, \cdots, D_K) \leq 3c(OPT)$

*Proof:* Assume the skyline of $OPT$ at $p_n$ is $D_1, \cdots, D_K$, we will prove that $c_n(D_1, \cdots, D_K) \leq 3w(OPT)$.

Since $OPT \in \mathcal{C}_n(D_1, \cdots, D_K)$, we have $c_n(OPT) \geq c_n(D_1, \cdots, D_K)$. Thus, we only need to prove that $c_n(OPT) \leq 3w(OPT)$. By Lemma 4, each disk can be counted at most 3 times in $c_n(OPT)$, thus the theorem holds. ∎

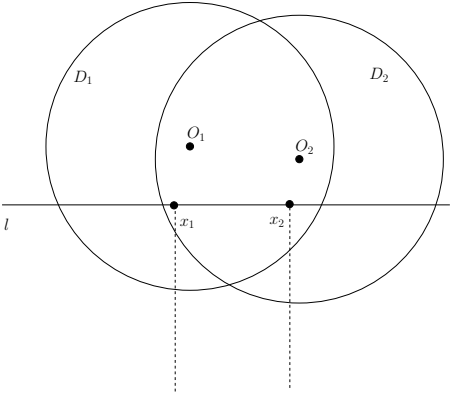

Fig. 3. The property of line-dominating.

The following lemma will serve as a technique basis for proving Theorem 4.

*Lemma 3:* [18] Consider two disks $D_1, D_2$ lying below above a horizontal line $l$, $D_1$ line-dominate $D_2$ w.r.t. $x = x_1$ and $D_2$ line-dominates $D_1$ w.r.t. $x = x_2$. Let $O_1, O_2$ be the centers of the disks $D_1, D_2$ respectively. Then $O_1$ lies to the left of $O_2$ iff $x_1 < x_2$ and vice versa. (Fig. 3).

*Lemma 4:* For any disk set $\mathcal{D}$ which is a $K$-cover of $\mathcal{P}$, each disk appears in the skylines non-consecutively for at most three times.

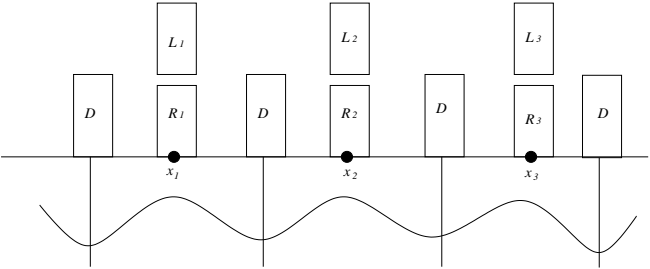

Fig. 4. Illustration for Lemma 4.

*Proof:* Assume to the contrary that some disk $D$ appears in the skylines non-consecutively for 4 times. Then, there exists three nodes $p_1, p_2, p_3$ (their $x$-coordinates are $x_1, x_2, x_3$ respectively) and $D$ appears in the skyline for three times: (1) before $x_1$ (2) between $x_1$ and $x_2$, (3) between $x_2$ and $x_3$, and (4) after $x_3$. However, $D$ does not appear in the skylines at $x_1$ or at $x_2$ or at $x_3$. We will derive contradiction. For the $K$ skyline-disks at $x_i : i = 1, 2, 3$, assume $L_i$ disks lie to the right of $D$ and $R_i$ disks lie to the right of $D$ Then, we have $L_1 + R_1 = L_2 + R_2 = L_3 + R_3 = K$. However, for any node before $x_1$, by Lemma 3, at least $L_1 + L_2 + L_3$ disks line-dominates $D$, we have $L_1 + L_2 + L_3 < K$ as $D$ appears in the skyline. For any node between $x_1$ and $x_2$, by Lemma 3, at least $R_1 + L_2 + L_3$ disks line-dominates $D$, we have $R_1 + L_2 + L_3 < K$ as $D$ appears in the skyline. For any node between $x_2$ and $x_3$, by Lemma 3, at least $R_1 + R_2 + L_3$ disks line-dominates $D$, we have $R_1 + R_2 + L_3 < K$ as $D$ appears in the skyline. For any node after $x_3$, by Lemma 3, at least $R_1 + R_2 + R_3$ disks line-dominates $D$, we have $R_1 + R_2 + R_3 < K$ as $D$ appears in the skyline. Thus, we have $(L_1 + L_2 + L_3) + (R_1 + L_2 + L_3) + (R_1 + R_2 + L_3) + (R_1 + R_2 + R_3) < 4K$. This means that $(L_1 + 3R_1) + 2(L_2 + R_2) + (3L_3 + R_3) < 4K$. However, we have $(L_1 + 3R_1) + 2(L_2 + R_2) + (3L_3 + R_3) = 4K + 2R_1 + 2L_3 > 4K$. This causes contradiction, which finishes the proof. ∎

## V. LITERATURE REVIEW

As we know, dominating set for uniform communication ranges is a classical geometric *set cover* problem. It is NP-hard [9] while there exist some constant-approximation algorithms for it, in contrast to the fact that the general set cover problem is not approximable within $O(\log n)$, where $n$ is the size of the input instance [14]. For a classical generalization of minimum dominating set, which is *min-weight dominating set*, a lot of recent results appeared in [1], [5], [6], [19]. the best result so far is achieved by [21] which proposed a $(4 + \epsilon)$-approximation algorithm in the setting of unit disk graph (UDG).

For the related disk cover problem, one branch that receives great research interest is called *discrete unit disk cover*, and there have been a series of work done for it [2]–[4], [13]. Specifically, Brönnimann and Goodrich [2] presented a deterministic $\epsilon$-net based algorithm where the constant factor is not specified. Calinescu *et al.* [3] gave a 102-approximation algorithm. Narayanappa and Vojtechovsky [13] improved the approximation ratio to 72. Carmi *et al.* [4] gave a 38-approximation algorithm by solving a subproblem where the points are located below a line and to be covered by a subset of disks of above the line.

Yun *et al.* [20] studied deployment patterns to achieve full coverage and $k$-connectivity under different ratios of the sensor communication range to the sensing range for homogeneous wireless networks. For the $k$-coverage problem where each target is required to be covered with multiplicity $k$, Wan *et al.* [17] analyzed the probability of the $k$-coverage when the sensing radius or the number of sensors changes while taking the boundary effect into account.

Recently, Mustafa and Ray [11], [12], proposed a PTAS for the discrete geometric hitting set problem. Based on their techniques, Gibson *et al.* [7], [8] gave a PTAS for the un-weighted case, and $2^{O(\log^* n)}$-approximation for the weighted case of the problem minimum dominating set in disk intersection graph with arbitrary disk radii.

## VI. CONCLUSION

Many challenging issues in wireless networks are intrinsically related to dominating set. In this work, we have presented several approximation algorithms for mutually related problems: minimum dominating set in disk containment graph, Strongly dominating set in directed disk containment graph, and linear $K$-Cover. For the problem linear $K$-Cover, this is the first time in the literature that the geometric properties for $K$-coverage with was explored, and the approximation algorithmic results was obtained. A possible future research direction is whether we can achieve constant approximation for the general $K$-Coverage problem based on the geometric properties.

## REFERENCES

[1] AMBUHL, C., ERLEBACH, T., MIHAL AK, M., AND NUNKESSER, M. Constant-Factor Approximation for Minimum-Weight (Connected) Dominating Sets in Unit Disk Graphs. *Lecture Notes in Computer Science 4110* (2006), 3.
[2] BR"ONNIMANN, H., AND GOODRICH, M. Almost optimal set covers in finite VC-dimension. *Discrete and Computational Geometry 14*, 1 (1995), 463–479.
[3] CĂLINESCU, G., MANDOIU, I., WAN, P., AND ZELIKOVSKY, A. Selecting forwarding neighbors in wireless ad hoc networks. *Mobile Networks and Applications 9*, 2 (2004), 101–111.
[4] CARMI, P., KATZ, M., AND LEV-TOV, N. Covering points by unit disks of fixed location. *Algorithms and Computation* (2007), 644–655.
[5] DAI, D., AND YU, C. A $5 + \epsilon$-approximation algorithm for minimum weighted dominating set in unit disk graph. *Theoretical Computer Science 410*, 8-10 (2009), 756–765.
[6] ERLEBACH, T., AND MIHALÁK, M. A $(4+ \varepsilon)$-Approximation for the Minimum-Weight Dominating Set Problem in Unit Disk Graphs. *Approximation and Online Algorithms* (2010), 135–146.
[7] GIBSON, M., AND PIRWANI, I. Algorithms for Dominating Set in Disk Graphs: Breaking the logn Barrier. *Algorithms–ESA 2010*, 243–254.
[8] GIBSON, M., AND PIRWANI, I. Approximation algorithms for dominating set in disk graphs. *Arxiv preprint arXiv:1004.3320* (2010).
[9] JOHNSON, D. The NP-completeness column: an ongoing guide. *Journal of algorithms 13*, 3 (1992), 502–524.
[10] MATULA, D., AND BECK, L. Smallest-last ordering and clustering and graph coloring algorithms. *Journal of the ACM (JACM) 30*, 3 (1983), 427.
[11] MUSTAFA, N., AND RAY, S. Improved results on geometric hitting set problems. *www. mpi-inf. mpg. de/˜ saurabh. Papers/Hitting-Sets. pdf* (2009).
[12] MUSTAFA, N., AND RAY, S. PTAS for geometric hitting set problems via local search. In *Proceedings of the 25th annual symposium on Computational geometry* (2009), ACM, pp. 17–22.
[13] NARAYANAPPA, S., AND VOJTECHOVSKỲ, P. An improved approximation factor for the unit disk covering problem. In *Proc. Can. Conf. on Comp. Geom* (2006), Citeseer.
[14] RAZ, R., AND SAFRA, S. A sub-constant error-probability low-degree test, and a sub-constant error-probability PCP characterization of NP. In *Proceedings of the twenty-ninth annual ACM symposium on Theory of computing* (1997), ACM, pp. 475–484.
[15] VARADARAJAN, K. Weighted geometric set cover via quasi-uniform sampling. In *Proceedings of the 42nd ACM symposium on Theory of computing* (2010), ACM, pp. 641–648.
[16] WAN, P., XU, X., AND WANG, Z. Wireless coverage with disparate ranges. In *Proceedings of the Twelfth ACM International Symposium on Mobile Ad Hoc Networking and Computing* (2011), ACM, p. 11.
[17] WAN, P., AND YI, C. Coverage by randomly deployed wireless sensor networks,. *IEEE Transactions on Information Theory, vol. 52, pp. 2658-2669* (2006).
[18] XU, X., TANG, S., MAO, X., AND LI, X. Distributed gateway placement for cost minimization in wireless mesh networks. In *Distributed Computing Systems (ICDCS), 2010 IEEE 30th International Conference on* (2010), IEEE, pp. 507–515.
[19] YAOCHUN HUANG, XIAOFENG GAO, Z. Z., AND WU, W. A better constant-factor approximation for weighted dominating set in unit disk graphs. *J.Comb.Optim. ISSN:1382-6905* (2008), 1573–2886.
[20] YUN, Z., BAI, X., XUAN, D., LAI, T., AND JIA, W. Optimal Deployment Patterns for Full Coverage and k-Connectivity ($k \leq 6$) Wireless Sensor Networks. *IEEE/ACM TRANSACTIONS ON NETWORKING 18*, 3 (2010).
[21] ZOU, F., WANG, Y., XU, X., LI, X., DU, H., WAN, P., AND WU, W. New approximations for minimum-weighted dominating sets and minimum-weighted connected dominating sets on unit disk graphs. *Theoretical Computer Science* (2009).

## APPENDIX

*Proof of Lemma 1:* To verify the first fact, consider a disk $A$, for any disk $D \in N(A)$, if $x_D \geq \frac{1}{2}$, we add $\lfloor 2n \cdot x_D \rfloor \geq n$ copies of $D$ to $\mathcal{D}_0$, then $A$ is at least $n$-dominated. Otherwise, we define $n$ disjoint subsets of disks, the $i$-th ($1 \leq i \leq n$) subset $\mathcal{D}_p^i$ contains every disk $D \in N(A)$ satisfying $\frac{i-1}{2n} \leq x_D < \frac{i}{2n}$. Since $x_D < \frac{1}{2}$ for each disk $D \in N(A)$, these $n$ subsets form a partition of $N(A)$. Let $x_1, x_2, \cdots, x_n$ denote the cardinalities of the $n$ subsets respectively. We have $\sum_{1 \leq i \leq n} x_i \cdot \frac{i}{2n} > \sum_{D \in N(A)} x_D \geq 1$, this means that $\sum_{1 \leq i \leq n} i \cdot x_i > 2n$. Since $\sum_{1 \leq i \leq n} x_i \leq n$, we have $\sum_{1 \leq i \leq n} (i-1) \cdot x_i > n$, thus $\sum_{D \in N(A)} \lfloor 2n \cdot x_D \rfloor \geq \sum_{1 \leq i \leq n} (i-1) \cdot x_i > n$. This implies that $\mathcal{D}_0$ at least $n$-dominate the disk $A$.

To verify the second fact, observe that $w(\mathcal{D}_0) = \sum_{D \in \mathcal{D}} \lfloor 2n \cdot x_D \rfloor < \sum_{D \in \mathcal{D}} 2n \cdot x_D = 2n \cdot \sum_{D \in \mathcal{D}} x_D = 2n \cdot \lambda^*$.

*Proof of Lemma 2:* The proof is similar to a similar result by Gibson *et al.* [8]:

**Lemma 5:** Let $\mathcal{D}$ be a set of $m$ disks, and $1 \leq L \leq m$ be an integer. Let $\mathcal{Q}$ be a set of disks (possibly infinite). There are $O(mL^2)$ disks of $\mathcal{Q}$ that intersect distinct subsets of $\mathcal{D}$, each of size at most $L$.

*Proof of Theorem 1:* By the proposed algorithm in Table II, the output $\mathcal{D}'$ at least $\log L$ dominate every equivalence class if this class is $L$-dominated in $\mathcal{D}$. We next show that $\Pr(D \in \mathcal{D}') \leq \frac{c \log L}{L}$. If a disk $D$ is not forced, clearly, $\Pr(D \in \mathcal{D}') = \frac{\log L}{L} \leq \frac{c \log L}{L}$. We are only left to upper bound the probability that a disk is forced.

For each disk $D_j$ in the sequence $\sigma$, If the disk $D_j$ is forced for some equivalence class, then all the disks $D_{j'}$ (with $j' \geq j$) that dominate this equivalence class are also forced, and the number of such disks is at most $\log L - 1$ (since otherwise $D_j$ will not be forced). At the same time, some disk $D_i$ ($i \leq j$) must be *first forced* for this equivalence class. Here a disk is first forced for an equivalence class if it appears as the first one among all forced disks for the equivalence class according to the order $\sigma$. We have the following inequality:

$\Pr(D_j$ is forced for an equivalence class$)$
$\leq \log L \cdot \Pr(D_i \ (i \leq j)$ is first forced for the equivalence class$)$

We next compute the probability of a disk $D_i$ being first forced for an equivalence class. Let $\mathcal{D}_{p,i}$ be the set of all disks in $\{D_k : k \leq i\}$ that dominate this equivalence class, and $n_1$ be the number of disks in $\mathcal{D}_{p,i}$ that is finally selected, then the following three facts hold:
(1) $D_i$ dominate the equivalence class $e$,
(2) $|\mathcal{D}_{e,i}| \geq L - \log L \geq \frac{L}{2}$,
(3) $n_1 \leq \log L$.

Clearly, $\Pr(D_i$ is first forced for equivalence class $)$ is at most the probability that fact (3) holds when both facts (1), (2)

are true, which means less than $\log L$ disks are finally selected for $\mathcal{D}_{e,i}$ while each disk from $\mathcal{D}_{e,i}$ is selected with probability at least $\frac{c \log L}{L}$. This can be reduced to a coin toss problem: in a sequence of at least $\frac{L}{2}$ coin tosses, each coin turns up head with probability of at least $\frac{c \cdot \log L}{L}$, less than $\log L$ coins turn up heads. Based on Chernoff bound, the probability is at most $\frac{1}{e^{\frac{c \log L}{16}}}$. Thus

$$\Pr(D_i \text{ is first forced for equivalence class } e) \leq \frac{1}{e^{\frac{c \log L}{16}}}$$

Since there are at most $2c'L^2$ classes of $C_j$ covered with multiplicity in $[L, 2L]$ and covered by $D_j$, $D_j$ can be forced for any one of the at most $2c'L^3$ equivalence classes. To sum up, we have the following inequalities:

$$\begin{aligned}
&\Pr(D_j \text{ is forced}) \\
\leq\ & 2c'L^3 \cdot \Pr(D_j \text{ is forced for an equivalence class}) \\
\leq\ & 2c'L^3 \cdot \log L \cdot \\
& \Pr(D_i(i \leq j) \text{ is first forced for the equivalence class}) \\
\leq\ & 2c'L^3 \cdot \log L \cdot \frac{1}{e^{\frac{c \log L}{16}}}
\end{aligned}$$

By setting appropriate values for the constants $c$ and $c'$, we can ensure that the probability of a given disk being selected is at most $\frac{c \log L}{L}$ in the proposed algorithm (in Table II). This finishes the proof.